# Minkowski and Special Relativity: Does His Spacetime Geometry Explain Space Contraction?

**Paul A. Klevgard, Ph.D.**

pklevgard@gmail.com

**Abstract**

For over a century Minkowski's spacetime has dominated discussions of space contraction and time dilation within special relativity. Brown and Pooley have called into question both Minkowski's assumptions and the effects his presumed spacetime has upon objects in motion. But while they reject Minkowski, Brown and Pooley do not fill in the missing causal connection between velocity and space contraction and time dilation. To supply this causal connection between object and observer in relative motion we should be focusing on energy difference rather than velocity difference. When different inertial observers at different relative velocities measure the same rod (or clock), each observer registers a different amount of kinetic energy for the object. Following the consequences of this permits a new understanding of relativistic space contraction and time dilation.

**Keywords:**     Minkowski, Brown, Pooley, relativity, spacetime, space contraction, time dilation, de Broglie

## 1.0     Introduction

The exact nature of Minkowski's spacetime continues to divide those who seek to understand special relativity. For some, spacetime is geometry made physical. Spacetime is often regarded as a substantival container which determines space and time partition, and therefore measurement readings, between inertial systems. But others are skeptical and for them spacetime is simply interesting mathematics and has no physical presence and no determinative (causal) effects. The skeptics find Minkowski's explanation for dimensional warpage (space contraction and time dilation) unconvincing and argue that spacetime has merely an inferred reality. This essay presents an explanation for space contraction and time dilation of moving objects that does not depend upon an inferred reality. For convenience in the pages that follow, inertial objects that move relative to an observer will be referred to as projectiles whatever their size.

Velocity (relative to an observer) and rest mass together produce space contraction and time dilation. Both rest mass and velocity must be present. If you have rest mass but no velocity relative to the observer then you have no dimensional warpage (space contraction, time dilation). The same is true if you have velocity but no rest mass which is the case with the photon. We can learn a lot about space contraction and time dilation by looking at where it does not occur. And so we begin with a review of some common characteristics of rest mass at zero velocity and the photon at maximum velocity.

**2.0  Field /Wave Form and Pure Entities**

Projectiles reside between two velocity endpoints. At the low end you have inertial matter that is stationary in space for some observer. At the high end you have radiation (the photon) that is stationary in time (by its own measure).[1] Although the two extremes are very different, they do bracket projectile motion and they do exhibit purity of mass or of energy. Space-stationary matter has rest mass but no kinetic energy. Time-stationary radiation has kinetic energy but no rest mass. It is convenient to describe inertial matter and radiation (the photon) as "**pure entities**." While matter in motion mixes rest mass and kinetic energy, pure entities do not.

Both pure entities (endpoints) are characterized by a *form* because something (mass or energy) is occupying a dimension. A form resides in a dimension but it is not separate from the matter or energy which gives that form reality in that dimension. Physicists confine themselves to entities of mass and energy, but those entities in turn possess a form.

Material objects extend in (occupy) space and we may characterize such objects as possessing the field-form. "Field" is used here in the simple sense as a space-volume descriptor whose characteristics (having a shape, not occurring) are derived from the existing object. Thus a material particle has a field form (a volume with a density measure) and so does the potential energy that surrounds a charged particle or an ion. A field may have a regular shape (e.g., a bocce ball) or an irregular shape (e.g., a chain of carbon atoms), but it must occupy that space either as mass (matter) or as potential energy.

The other pure entity, radiation (the photon), possesses the waveform at least until it terminates.[2] Just as matter with its field form occupies space, so the photon with its waveform extends in ("occupies") time since radiation oscillation requires a time interval.

Pure entities do not mix rest mass and kinetic energy. They are either space-stationary matter or time-stationary radiation. They share certain formal attributes at a foundational (ontological) level: they are quantized, each has a form and each occupies an interval in one dimension (space or time) while progressing in the other dimension. But there are differences between our pure entities that follow from the fact that one exists and the other occurs.

Rest mass (matter) exists; it persists over time. You can convert mass (via $E = mc^2$) but you cannot destroy it. Rest mass is also frame independent; it is objective for all inertial observers whatever their relative velocity. On the other hand kinetic energy is relational, it involves relative velocity and therefore it is frame dependent. Kinetic energy can be increased, decreased or even eliminated by the choice of inertial frames.[3] Of course the phrase "eliminating kinetic energy by a choice of inertial frames" is too simplistic a statement. In the real world an observer can only 'eliminate' the kinetic energy (relative velocity) of an observed object (projectile) by catching up to the inertial frame of that object and that means the observer has acquired energy (acceleration over space equals work done). "Eliminating" kinetic energy is therefore a relative thing involving energy acquired by the observer's reference frame.

---

[1] Discussed in Section 5.0.
[2] The photon can interfere with itself making it a wave. It does terminate discretely simulating particle impact. But photons pass through each other as waves do; mass particles impact each other. The photon as "particle" requires a different essay.

So kinetic energy whatever its variety, radiation or projectile motion, is always frame dependent unlike rest mass. This reminds us that existence with its field form and occurrence with its waveform are ontologically orthogonal even as they come together as matter-in-motion. The joining of one form with another needs to be spelled out.

### 3.0   Projectiles as a Union of Mass and Energy

Projectiles approaching either pure entity resemble that entity in form.[4] Slow moving, large projectiles resemble the field-form stationary material object; small projectiles such as electrons approaching the speed of light display de Broglie waves and thereby resemble the photon in (wave) form.[5] The kinetic energy of projectiles manifests itself as relativistic mass via the equation $m_{rel} = m_{rest}(\gamma - 1)$ where $\gamma = 1/\sqrt{1 - v^2/c^2}$. The ratio of relativistic mass to rest mass is therefore a measure of the kinetic energy of a projectile. The waveform of very fast projectiles is a consequence of their relativistic mass being enormous compared to their rest mass. The field form of slow projectiles is a consequence of their rest mass dominating their relativistic mass.

- **Projectiles are a combination of rest mass and relativistic mass (the latter due to kinetic energy). The relative proportion of one to the other determines whether field-form or waveform dominates.**

The first laboratory confirmation of de Broglie radiation (matter waves[6]) came in 1926-27 with the Davisson-Germer experiments on diffracting electrons. Wave behavior then (and now) became inextricably linked with diffraction effects. But the idea that only particles of tiny mass (and high energy) have a wave character is quite wrong. Of course the larger the projectile's mass the shorter its wavelength since an object's momentum is inversely proportional to the wavelength $\lambda$ of an object: $\lambda = h/mv$ where '$m$' is total mass, rest mass plus relativistic mass. This means that large objects at high velocity (meter sticks, spaceships) can never diffract. But they still generate the same de Broglie wave character as a much smaller mass at the same velocity. The wave identity of a high speed projectile is always present regardless of projectile mass.[7]

---

[3] This is true for the kinetic energy of radiation just as it is for the kinetic energy of projectiles. Moving toward (away) from a light source increases (decreases) the light's frequency and hence its energy. If you could move away from the source at the speed of light the light's frequency and kinetic energy would be zero.

[4] This is a variant of Bohr's correspondence principle.

[5] Electron waves are not solely a consequence of linear velocity. "It is found that an electron which seems to us to be moving slowly, must actually have a very high frequency oscillatory motion of small amplitude superposed on the regular motion which appears to us. As a result of this oscillatory motion, the velocity of the electron at any time equals the velocity of light." P.A.M. Dirac Nobel Address of 1933.

[6] "Energy waves" is a more accurate term than "matter waves." There are no waves without the kinetic energy of motion.

### 3.1    Entities Again

We now have three types of entities based upon mass and energy. There is existing (inertial) mass that is free of (kinetic) energy; there is radiation energy that is free of (rest) mass; and there is the combination of rest mass with kinetic energy (a projectile).

An entity must always have a form; inertial mass has the field form, radiation energy has the waveform and the projectile has both field form and waveform. The projectile's form depends upon the projectile's kinetic energy, and hence its velocity, relative to an observer. A projectile, regardless of size or mass, moving VERY fast relative to an inertial observer is mostly waveform since relativistic mass dominates rest mass. That same projectile moving slowly relative to another inertial observer is mostly field form. So two (or more) inertial observers cannot agree on the wave/field form of the projectile; they also cannot agree on the projectile's extension (length) in space nor upon its progression in time. Are differences in form related to differences in space and time measures?

### 4.0    Space Contraction

We have seen (Section 2.0) that the field form extends in (occupies) space whereas the waveform does not. For entities possessing those forms, mass occupies space and radiation does not. Radiation may be detected in a space interval, but no wave can exclude a similar wave from that same interval.

Like waves can *superimpose* on each other and occupy no space in doing so. This is true of waves of any variety: water waves, sonic waves, electromagnetic waves and de Broglie waves. There is no limit to the number of electromagnetic waves within a black box cavity just as there is no limit to the number of de Broglie waves that create a wave packet. Waves cannot possibly add to the space a projectile's rest mass occupies, but can they reduce that space? This comes down to the model we choose for the interaction of waveform and field form in projectile motion.

Suppose the field form of the projectile's rest mass stays separate from the projectile's de Broglie waveform. Each form (field, wave) and each corresponding constituent (rest mass and de Broglie wave respectively) will then relate to space as if the other was not present. The rest mass with its field form will not be compromised; it will extend in space just as if it were stationary there. In this case rest mass extension is not contracted due to the presence of a waveform. This is like raisins in a pudding;[8] each constituent keeps its own identity.

On the other hand, suppose that a projectile's field form (from rest mass) and its waveform (from de Broglie waves) blend-fuse together into something intermediate. This is like field-form red wine combining with waveform white wine to create rosé wine. In this case the field form and its space extension are compromised by dilution with a waveform that occupies no space. So which is it? Raisin pudding or rosé wine?

---

[7] The wave packet of the large projectile is well-defined (compact) in space; not so for the very small projectile. This leads to the erroneous assumption that the large projectile does not have a wave packet.
[8] Plum pudding for the Brits. Cf. J.J. Thomson's atomic model of 1904.

**Raisin pudding:** **The case against the waveform causing space contraction** – The mass (matter) of a projectile has the space-occupying field form. This form and (rest) mass are unaffected and undiminished by the addition of de Broglie waveform that results from kinetic energy (relative velocity). Mass exists and is conserved; kinetic energy cannot undermine that. We have more than a century of tradition supporting the contention that the addition of kinetic energy does not alter rest mass; no empirical evidence has emerged to contradict that.

**Rose wine:** **The case for the waveform causing space contraction** – A projectile's mass and kinetic energy constitute a unified whole. The quantitative fusion of the two is expressed in the projectile's effective mass (rest mass plus relativistic mass). The merging of wave and field (i.e., wave and particle[9]) is found throughout physics. Material media host waves and waves or vibrations can reinforce to create field-like entities (e.g., the wave packet, standing waves, phonons, etc.). The Heisenberg uncertainty relation itself expresses the measurement limitations when waveform energy combines and blends with field form mass; precise measurements are impossible because wave and field have fused together. The two ontological forms, wave and field, are not exclusionary; in most of physics they are blurred together. We can see this in water waves but been able to ignore this with projectiles because we live in a world of large and massy objects possessing low energy and low velocity and de Broglie waves are not visible.

Raisin pudding is certainly the easier choice for the reader since it conforms to tradition and to our everyday experience where kinetic energy is a formless quantity. But it doesn't accord with how wave and field interact in the rest of physics. And it also leaves us stuck with the current, somewhat dubious, explanations (below) for space contraction and time dilation. The author opts for rosé wine.

Adding kinetic energy to rest mass creates a new, hybrid entity with its own identity, its own effective (inertial) mass and its own unique blend of existing rest mass field form and occurring de Broglie waveform. The blended constituents of the hybrid entity necessarily constitute 100 percent of the 'new' identity. Any increase of kinetic energy (velocity) and its waveform comes at the <u>percentage</u> decrease (not a quantitative loss) of rest mass and its field form. It is a zero sum game between wave and field; the percentage gain of one is a corresponding diminution for the other. And since the field form of rest mass occupies space whereas the waveform of superimposing de Broglie radiation does not, any increase of projectile velocity-and-waveform is accompanied by space contraction. As a single blended entity, the projectile occupies space to the degree that rest mass and its field form participate in the unified entity. Maximum space is occupied at zero relative velocity when there is no wave participation; minimum space is occupied very close to the speed of light when rest mass (of whatever quantity) is totally dominated by enormous kinetic energy.

---

[9] The phrase 'wave-particle' is empirically based and is a comparison of apples with oranges. A wave is a form; a particle is an entity. The proper dichotomy is between forms: wave versus field.

Wave character is independent of wave length which means that at the same velocity heavy projectiles with very small wavelengths spatially contract to the same extent as very light projectiles with larger wavelengths. Of course, comparing a projectile's wave character to its field character is a strictly qualitative approach. For a quantitative measure it is the ratio of relativistic mass to rest mass that tracks the change of space and time for matter in motion. For space contraction this ratio is the familiar Lorentz factor, $\sqrt{1 - v^2/c^2}$, which is near 1 at low velocities but is very small at velocities close to that of light (space has contracted). The ratio for time dilation is the reciprocal, $1/\sqrt{1 - v^2/c^2}$, which is also nearly 1 at low velocities (time is fast) but gets very large at values near the speed of light (time has slowed).

**5.0     Time Dilation**

Section 2.0 has argued that an inertial rest mass is stationary in space – zero space progression – by its own measure. That is true by the very definition of what constitutes an inertial state for a force-free material object. The equivalent statement for the photon – that it undergoes zero time progression by its own measure – is less obvious. Some who dispute this try to restrict the passage of time either to inertial frames (which the photon lacks) or to proper time (the time a clock would show when moving from point A to point B). This view is mistaken; photons are stationary in time as Andrew Holster (on the web, undated) points out.

> *"…the faster a system travels through space, the slower its internal processes go. At the maximum possible speed, the speed of light, c, the internal processes in a physical system would stop completely. Indeed, for light itself, the rate of proper time is zero: there is no 'internal process' occurring in light. It is as if light is 'frozen' in a specific internal state.*
> *… [I]n quantum mechanics, the most fundamental particles have an intrinsic proper time, represented by an internal frequency. This is directly related to the wave-like nature of quantum particles.*
> *For light, treated as a quantum mechanical particle (the photon), the rate of proper time is zero, and this is because it has no mass. But for quantum mechanical particles with mass, there is always a finite 'intrinsic' proper time rate, represented by the 'phase' of the quantum wave."*

De Broglie radiation can form an image in an electron microscope just as electromagnetic radiation can form an image in an optical microscope. In general we can say that radiation imaging is based upon waveform kinetic energy probabilistically impinging upon a target as discrete action quanta. Although electrons and photons are different quanta and have different energy sources (matter-in-motion versus electromagnetic oscillation), they still constitute radiation. And one (electromagnetic) is the upper limiting case of the other in terms of velocity and the dominance of waveform over field form as Holster points out. Radiation should be thought of generically as waveform kinetic energy that comes in two (quanta) flavors, de Broglie radiation and electromagnetic radiation.[10]

Electromagnetic radiation is frozen in time. If we could incorporate that radiation waveform into rest mass surely that would slow down the advance of time for that rest mass. Nature doesn't allow that, but does provide for the blending of rest mass with de Broglie radiation waveform and this does result in time dilation. Rest mass alone races along in time while (generic) radiation is stationary there; combining rest mass with radiation leads to time dilation just as it leads to space contraction.

Two observers in relative motion find the other to possess rods that are shortened and clocks that are slowed. Yet each observer finds his or her own rods and clocks to be perfectly normal. This is a consequence of the kinetic energy of moving objects to be relative to one's spatial reference (frame dependence, see Section 2.0). Each observer sees the other as possessing a velocity, a relativistic mass, a de Broglie waveform and a space contraction plus time dilation that s/he does not possess.

### 6.0    Executive Summary:

*A projectile fuses field form and (de Broglie) waveform into a single, hybrid (blended) form. Field form occupies space, waveform does not (waves superimpose in space – see Section 4.0). A projectile reflects its form and will space-contract to the degree that its hybrid form incorporates superimposing waveform. The photon as complete waveform has ultimate superposition-contraction and ultimate dilation; it occupies no space and its time is stationary.*

*The agent of change is the waveform of kinetic energy, not velocity itself.*

### 7.0    Lorentz and Minkowski

Both Lorentz and Minkowski offered explanations for dimensional warpage. For Lorentz the elementary constituents of reality were physical objects in space (including their electrostatic fields) which existed in time. The passage of such objects through the stationary aether affected their electronic binding at the atomic level resulting in space contraction. This explanation struck many as very reasonable because change was then a consequence of one material entity acting upon another.

---

[10] For the reader who finds no relationship between matter waves and photon waves, consider the following. De Broglie's formula, $p = h/\lambda$, has momentum as a measure of action $h$ over (divided by) cycle space (wavelength) and it applies to the photon as well as to matter-in-motion. The Planck – Einstein formula, $E = hf = h/T$, has energy as a measure of action $h$ over (divided by) cycle time (period T). Both forms of radiation have essentially the same function, the "delivery" of action packets in a space or a time interval. Both electromagnetic waves and matter waves (for a single particle/object) are in three-dimensional space. And de Broglie's derivation of matter waves relies upon the Lorentz transforms which presuppose a constant value for the velocity of light.

Unfortunately the aether remained undetectable; it was an inferred entity and it gradually fell out of favor. It became a victim of Ockham's razor.

Minkowski took a different approach. As a theoretical mathematician he was not committed to a traditional concept of physical reality such as material objects existing in time. For Minkowski the elementary constituents of reality were events that were based upon a fusion of space location of an object with its time location. In place of unchanging objects over time per Lorentz, Minkowski found reality to reside in an object's history taken as "the set of happenings that constitutes its life" (Sklar, 1974, p. 163). Minkowski's "world-points" resided in spacetime (a substantival container for some) both of which had an inferred reality. This redefinition of reality meant that measurements of space taken separately from time were mere artifacts; "real" measurements were intervals between events in spacetime and they were invariant for all inertial observers regardless of their relative velocities.[11]

These two theories of dimensional warpage certainly reflect that period in physics which produced them. The Lorentz theory came at the very end of an era when the aether was still considered necessary for the propagation of electromagnetic waves. Why not make the aether an agent of change for moving rods and clocks? The Minkowski theory was an early instance of the enthusiasm theorists developed for the mathematization of physics.[12]

Both Lorentz and Minkowski agreed that space contraction and time dilation were real for an observer. That is, dimensional warpage of rods and clocks moving relative to an observer was an objective and repeatable measure providing the measurement was done within the inertial system of the observer. But different inertial observers – each having a different relative velocity – can all measure the same rod over the same time span and find it to have different lengths. Depending upon an observer's direction of motion relative to the rod, the space contraction may affect the rod's length or its width. How can the same rod have different measurable dimensions at the same time?

As we have seen, Minkowski "solves" this puzzle in a very direct way with his grandly-named *"The Postulate of the Absolute World."* He argues that only spacetime intervals have reality because all observers can agree upon them. Space measures and time measures by themselves have no fundamental reality since they differ from observer to observer.

If multiple moving observers measure the "same" rod with different results, and if they are careful in those measurements, then only two possible conclusions follow. Option 1: the measured quantity, space, is at fault. That is, space itself is velocity-relative and therefore unreal or at least suspect as a measurement. Option 2: the observers are mistaken in their belief that they are measuring the same rod since different measurements imply different rods or at least different identities of the same rod. For Minkowski there was only Option 1 which he pursued with a great deal of confidence and zeal. But what for him was the only option was in fact the wrong option. Nature is much more subtle than he realized. As we have seen, observers in relative motion are **not** measuring the same rod identity because each sees the rod's blend of rest mass and kinetic energy differently.

---

[11] Invariance between different observers became much more important for later thinkers than Minkowski's notion that event-points constituted reality and that our local space metric was an illusion. Invariance seemed the one certainty to embrace for later Minkowski adherents. See Janssen (2008, p. 70).
[12] Theorists such as Minkowski's close friend, David Hilbert. See Schirrmacher, 2003 for a discussion of the trend toward mathematization.

Rods and clocks contract and dilate respectively for any inertial observer in relative motion because those objects have, for that observer, added kinetic energy with its waveform identity to a preexisting rest mass with its field form identity. Rods and clocks are an observer-relative, objective union of field-form rest mass and waveform kinetic energy. The result is a mixed projectile entity having a single blended form that determines entity space extension and entity time progression.

**8.0    Brown and Pooley and Their Critics**

Minkowski's treatment of special relativity as a mathematized geometry of four dimensions proved to be very successful. It remains as perhaps the best way of conceptualizing the subject. But whether it was the best, or only, way to explain why rods shrink and clocks run slow was open to question. Familiar to all who follow this subject, Harvey Brown took up this challenge of explanation in the 1990s and produced a landmark book on the subject in 2005. In it he writes (p. 23): "…if it is the structure of the background spacetime that accounts for… [space contraction and time dilation], by what mechanism is the rod or clock informed as to what this structure is?" Brown's critique raises the issue of causation forcing various scholars to redefine (or clarify, or deny) causation within Minkowski's spacetime.[13]

One response is to argue that any change (contraction) that varies from one observer to another is merely apparent change that does not require a causal explanation.[14] This overlooks the fact that some relativistic changes are objective for all observers. In the twin paradox where Bob remains on the earth and Alice journeys (rapidly!) to and from a distant star, Alice returns to earth younger than Bob and remains so. And this age difference between the birth twins on earth is observer-independent; it has the same value for any observer moving relative to the earth at any speed.

Another way to answer Brown's critique is to redefine causation. Dorato and Felline (2010, p. 3) argue for a geometric/structural explanation (of contraction, dilation) that is "independent of…attempts at explaining phenomena by invoking causal or mechanistic models." Such a structural explanation has no "…need [of] forces to account for the relativistic phenomena of contractions and dilations…" (p. 2).

Janssen (2008, p. 69) takes a similar approach by insisting that space and time deformations are part of kinematics and therefore they "are simply examples of the default or generic spatio-temporal behavior posited by the theory [of special relativity]." Minkowski spacetime is not a container that causes space and time warpage,[15] rather this warpage is "default behavior…encoded in the geometry of Minkowski space-time" (p. 69) and no further explanation is required. Janssen wishes to define kinematics to include those effects (contraction) that Brown regards as dynamic. Janssen writes (p. 63, his italics): "*if an effect can be defined away by a mere change of convention about how to slice Minkowski space-time, then that effect is purely kinematical.*"[16]

---

[13] "…the Minkowskian metric is no more than a codification of the behaviour of rods and clocks…" (Brown, 2005, p. 9). See Pooley (2012) for a discussion of the controversy and a bibliography.

[14] "Since one can talk about a definite amount of a contraction only relatively to an arbitrarily chosen inertial frame, and since by changing frame we change also the amount of contraction, there is no invariant/objective fact (the contraction) to be explained causally/dynamically." Dorato, 2014, p. 18. Janssen might agree but see note 16.

[15] Janssen even accepts the Brown and Pooley description of spacetime as a "non-entity." See (2008, p. 5).

Robert DiSalle also defends Minkowski against Brown and Pooley. He writes (2006, p. 116-117) that

> "Minkowski's account…reveals just how simple and direct is the connection between the structures of space and time and the assumptions we make about physics. The claim at the heart of Minkowski's analysis is…that a world in which special relativity is true, simply, *is* [sic] a world with a particular space-time structure."

Elsewhere DiSalle (2009, p. 2), like Janssen and Dorato, sidesteps causation when he writes "Minkowski's space-time explains the significance of Einstein's theory for our understanding of the world, and of the nature of space-time; it explicates the conceptual revision that Einstein's theory forces upon us." It is not clear from this whether it is the world or whether it is Einstein's theory that forces Minkowski spacetime upon us. DiSalle does hold the door (barely) open for a dynamical explanation (of contraction). He writes (2009, p. 2) that a "demand for a dynamical explanation…require[s] a physical interpretation that has yet to be adequately defined."

This is a limited sample and apologies to the authors for random excerpts, but none of these esteemed critics wants to go back to spacetime as an existing causal container. Brown and Pooley and Ockham (his razor) made that option unattractive. The critics all dispute the contention of Brown (and Einstein) that we need a constructive theory to explain dimensional warpage. The shared tendency of the critics is to redefine the meaning of kinematics or of causation or promote the efficacy of a vague geometric/structure. The common thread for all is that we do not need an "explanation" for dimensional warpage, or if we do we already have an unrecognized explanation. The many pages and all the ink devoted to helping us recognize what isn't obvious makes one yearn for an exposition that is less subtle, less nuanced and less prolix.[17]

## 9.0     Conclusions

Minkowski gets full credit for weaving ideas from his predecessors – Lorentz, Poincare , Einstein – into an elegant presentation of relativity in four dimensions. It remains a contribution of enduring value especially for visualizing special relativity. But his ontology of world points and an existing spacetime has fared less well over the decades. He was too close in time to the inception of special relativity to get everything right. His framework does not explain space contraction and time dilation.

---

[16] But a real-world "effect" cannot be eliminated merely by changing the angle to slice a geometric representation. If a speeding rod is space contracted by your measure (in your inertial system), you can only eliminate this contraction by 1) 'hopping aboard' the rod or 2) slowing the rod down to make it stationary in your inertial system. Either way you subtract kinetic energy from the rod by making it space stationary relative to you. Minkowskian geometry trivializes dimensional changes by hiding the change of kinetic energy. Change is not free.

[17] As Brown points out (2005, p. 158), once you give up causation (to explain dimensional warpage) and embrace vaguer concepts (structural background, default behavior, encoded in geometry, etc.) you can wind up with "another analogue of Molière's dormative virtue."

**Dimensional Warpage:** All of the force-free <u>entities</u> that are of interest to physicists – space-stationary matter and its fields (of charge, of gravitation), time-stationary radiation (photons), matter in motion – have but a single form. Two of these three entities have a purity of form: pure field-form for inertial matter or pure waveform for the photon. But for material objects in motion their single form is a blend of field form and wave form. The characteristics of field form (i.e., occupy space) will be overlaid and diminished when combined with the opposing characteristics of wave form (i.e., superposition in space). When mass and energy combine to create a unique, observer-relative entity (projectile), it is the mass-as-field and energy-as-wave balance that determines dimensional extension and progression. Space contraction and time dilation of material objects (rods, clocks) in motion are direct consequences of the addition of waveform energy to field-form matter.

If you take the equality of mass and energy seriously, it makes perfect sense that: 1) mass should exist and have one existing form; 2) radiation energy should occur and have a different occurring form; 3) each may be found in their pure form; and 4) the two of them may combine as one blended form.

**The Weight of Tradition**: It is difficult to give up familiar concepts and explanations. In physics it is often new experimental results that force the issue as when the Davisson and Germer confirmed the rather dubious claims of de Broglie about matter waves. But sometimes there are two or more ways to explain a phenomenon and there is no decisive experiment to guide our choice between them. That is the case with space contraction and time dilation. Minkowski's explanation gained textbook status over the decades and became a soft pillow upon which to sleep, to borrow from Einstein. But Brown and Pooley have interrupted that sleep. It seems to be a good time for readers with open minds to consider new ideas on the subject.

This essay has explained how the observer-relative kinetic energy of a material object can account for dimensional change by altering the form of an object. This explanation offers a number of advantages over other theories.

- First, it does not depend upon any inferred entity functioning either as a container or as an agent of change.
- Second, it provides a causation that is intrinsic to a material object and its form; warpage is unique for each observer and objective only within that observer's inertial system.
- Third, it is a constructive theory and not a principle theory; that is, explanation begins at a foundational level.[18]
- And finally, as a constructive theory it bypasses the messy issues of molecular binding (or fictitious forces) when considering a rod's space contraction and the internal strain that implies.

***To each unique observer a unique rod (and clock) is given, each with a distinct, blended form.***

---

[18] For the distinction see Brown & Pooley (2004, section 3). Also Balashov & Janssen (2003, p. 331).

**Appendix A – The Twin's Paradox**

One of the most famous paradoxes of special relativity involves the thought experiment of Bob and his twin sister Alice. Bob stays on earth while Alice travels to and back from a distant star at eight-tenths of the speed of light. While in flight Alice is space contracted and time dilated according to Bob. When Alice finally returns to earth (to Bob's inertial system) she has only aged 10 years while Bob has aged 16.6 years. But Alice is the same size as when she left; space contraction has been reversed, but the effects of time dilation remain.

Actual experiments have been conducted with identical, synchronized atomic clocks, one "at rest" on the earth (and therefore rotating inertially) and one subjected to repeated flights on jet planes. When the flights are over and the clocks are finally compared side-by-side, the travelling clock has lost time but its size in space is identical to its twin.

This lack of symmetry is a consequence of how dimensional warpage affects progression differently than it does extension. Alice extends in (occupies) space and her space contraction in flight gets reversed when her flight is over. Extension is like a quantity: that which is taken away during flight gets restored when the flight concludes.

But things are different for the dimension wherein Alice progresses, namely time. Her progression rate in time slows during flight and her rate will speed back up to Bob's rate when her flight is over. But the aging interval that Alice lost due to her temporal slowdown is permanent. There is no way Alice can recover the time interval lost when her aging (progressing) rate fell behind that of Bob.

While field-form material entities such as Alice progress in time, waveform energy entities (the photon) progress in space. Hence a photon can have a "lost space" interval just as Alice experiences a "lost time" interval. There are two ways a photon can have a lost space interval, one more practical than the other.

The obvious, practical way is for the photon to travel in a vacuum, enter a medium such as glass or water, and then emerge back into vacuum. Although the emerging photon resumes its vacuum speed, it has incurred a lost space interval due to its reduced velocity in the transparent medium.

The less practical way for the photon to have a lost space interval is to duplicate Alice's experience but in reverse. The scenario for Alice is as follows. She is rest mass that acquires kinetic energy (relative velocity) and has a lost time interval after giving up her kinetic energy (her velocity) and returning to the inertial system of her twin brother.

The scenario for the photon must involve the adding and subtracting of rest mass (matter) rather than kinetic energy. Somehow the kinetic energy photon acquires a small amount of rest mass (matter) and thereby slows its rate of space progression. After this "heavy" photon sheds its rest mass and resumes regular photon velocity it has a lost space interval it can never make up. While photons in practice may not behave this way, that is not the point. The point is that lost intervals always belong to the progression dimension. For rest mass (matter, namely Alice) the interval lost is in time; for kinetic energy (the photon) the interval lost is in space.

If you accept the idea that entities extend in one dimension (occupy an interval or volume) and progress in another, then this explanation makes sense. If, like Minkowski, you reject any distinction between extension and progression dimensions, then this explanation makes no sense at all.

_______________________________________________